\documentclass[a4paper,12pt]{article}
\usepackage{mathrsfs} \usepackage{amsmath} \usepackage{amssymb} \usepackage{slashed}
\usepackage{graphicx} \usepackage{caption}
\begin{document}
\newcommand{\ba}{\begin{eqnarray}} \newcommand{\ea}{\end{eqnarray}}
\newcommand{\be}{\begin{equation}} \newcommand{\ee}{\end{equation}}
\renewcommand{\figurename}{Figure}
\renewcommand{\thefootnote}{\fnsymbol{footnote}}

\vspace*{1cm}
\begin{center}
 {\Large\textbf{Neutrino Mass, Leptogenesis, and Dark Matter from The Dark Sector with $U(1)_{D}$}}

 \vspace{1cm}
 \textbf{Wei-Min Yang}

 \vspace{0.3cm}
 \emph{Department of Modern Physics, University of Science and Technology of China, Hefei 230026, P. R. China}

 \emph{E-mail: wmyang@ustc.edu.cn}
\end{center}

\vspace{1cm}
\noindent\textbf{Abstract}: I suggest a new extension of the SM by introducing a dark sector which has several new particles and a local $U(1)_{D}$ symmetry. The dark particles bring about the new and interesting physics beyond the SM. The model can generate the tiny neutrino mass by a hybrid see-saw mechanism, achieve the leptogenesis at the TeV scale, and account for the cold dark matter. All of the three things collectively arise from the dark sector. In particular, it is very feasible to test the model predictions and probe the dark sector in near future experiments.

\vspace{1cm}

\newpage
\noindent\textbf{I. Introduction}

\vspace{0.3cm}
  The standard model (SM) of the particle physics has successfully accounted for all kinds of the physics at or below the electroweak scale, refer to the reviews in Particle Data Group \cite{1}, but it can not explain the three important issues: the tiny neutrino mass \cite{2}, the matter-antimatter asymmetry \cite{3}, and the cold dark matter (CDM) \cite{4}. Many theories have been suggested to solve these problems. The tiny neutrino mass can be generated by the seesaw mechanism \cite{5} or origin from the loop-diagram radiative generation \cite{6}. The baryon asymmetry can be achieved by the thermal leptogenesis \cite{7} or the electroweak baryogenesis \cite{8}. The CDM candidates are possibly the sterile neutrino \cite{9}, the lightest supersymmetric particle \cite{10}, the axion \cite{11}, and so on. These theories ordinarily focus on only one of the three problems. In recent years, some inspired ideas attempt to find some connections among the neutrino mass, the baryon asymmetry, and the CDM, for example, the lepton number violation at the super-high scale can lead to the neutrino mass and the leptogenesis \cite{12}, the neutrino mass and the leptogenesis can also be implemented by the super-heavy scalar triplet \cite{13}, the asymmetric CDM is related to the baryon asymmetry \cite{14}, and some sophisticated models unifying them into a frame \cite{15}. Although many progresses on these fields have been made all the time, an universal and convincing theory is not established as yet.

  What is exactly a realistic theory beyond the SM? The universe harmony and the nature unification are a common belief of mankind. It is very possible that there is a common origin of the tiny neutrino mass, the matter-antimatter asymmetry and the CDM, which relates the three things to each other. Therefore, a new theory beyond the SM should be capable of accounting for the three things collectively, on the other hand, this theory should keep such principles as the simplicity, the fewer number of parameters, and being feasible and promising to be tested in future experiments. If one theory is excessive complexity and unable to be tested, it is unbelievable and infeasible. Based on these considerations, I suggest a new extension of the SM. The model introduces a dark sector beyond the SM, which contains a few of new particles and has a local gauge symmetry of $U(1)_{D}$, in particular, the dark sector provides a common origin of the above three things so that the model can completely account for them. Finally, it is very feasible to test the model and probe the dark sector by the TeV-scale colliders, the underground detectors, and the cosmic ray search.

  The remainder of this paper is organized as follows. I outline the model in Section II. Section III and Section IV are respectively discussions of the leptogenesis and the dark matter. Section V is  numerical results and discussions of the model test. Section VI is devoted to conclusions.

\vspace{0.6cm}
\noindent\textbf{II. Model}

\vspace{0.3cm}
  The model introduces several new particles and a local gauge symmetry $U(1)_{D}$ beyond the SM sector, which are in the dark sector. In addition, it obeys the global symmetry $U(1)_{B-L}$, i.e., the difference between the baryon number and the lepton one is conserved. Table. 1 clearly lists the particle contents and their quantum numbers of the model, all kinds of the notations are self-explanatory.
\begin{alignat}{1}
&\hspace{1.2cm}\begin{array}{|c|l|l|}
 \hline &\hspace{1.0cm}\mbox{SM sector} &\hspace{1.2cm}\mbox{Dark sector}\\
 \hline &\mbox{Higgs}\, \vline \hspace{0.6cm}\mbox{Lepton}&\hspace{0.2cm}\mbox{Fermion} \hspace{0.3cm}\vline \hspace{0.8cm}\mbox{Scalar}\\
 \hline &\hspace{0.25cm}H\hspace{0.445cm}\vline \hspace{0.2cm} l_{L}\hspace{0.45cm}e_{R}\hspace{0.4cm}N_{R} &\hspace{0.3cm}N_{L}\hspace{0.45cm}\chi_{R}\hspace{0.17cm} \vline \hspace{0.3cm}\phi_{1}\hspace{0.5cm}\phi_{2}\hspace{0.5cm}\Phi\\
 \hline SU(2)_{L} &\hspace{0.3cm}2\hspace{0.8cm}2\hspace{0.65cm}1\hspace{0.65cm}1 &\hspace{0.4cm}1\hspace{0.75cm}1\hspace{0.8cm}1\hspace{0.7cm}1\hspace{0.68cm}2\\
 \hline U(1)_{Y} &-1\hspace{0.28cm}-1\hspace{0.1cm}-2\hspace{0.65cm}0 &\hspace{0.4cm}0\hspace{0.75cm}0\hspace{0.8cm}0\hspace{0.7cm}0\hspace{0.17cm}-1\\
 \hline U(1)_{B-L}^{global} &\hspace{0.3cm}0\hspace{0.3cm}-1\hspace{0.12cm}-1\hspace{0.14cm}-1 &\hspace{0.4cm}0\hspace{0.75cm}0\hspace{0.8cm}1\hspace{0.7cm}0\hspace{0.18cm}-1\\
 \hline U(1)_{D} &\hspace{0.3cm}0\hspace{0.8cm}0\hspace{0.62cm}0\hspace{0.68cm}0 &\hspace{0.4cm}1\hspace{0.75cm}1\hspace{0.8cm}1\hspace{0.7cm}2\hspace{0.68cm}1\\
 \hline Z_{2} &-1\hspace{0.25cm}-1\hspace{0.65cm}1\hspace{0.65cm}1 &\hspace{0.07cm}-1\hspace{0.75cm}1\hspace{0.28cm}-1\hspace{0.72cm}1\hspace{0.66cm}1\\
 \hline
\end{array}
\nonumber\\
&\mbox{Table 1. The particle contents and the quantum numbers of the model.}\nonumber
\end{alignat}
Here I omit the quark sector and the color subgroup $SU(3)_{C}$ since what followed will not involve them. Each of the fermions in Tab. 1 has three generations as usual. $l_{L}=(\nu_{L}^{0},e_{L}^{-})^{T}$, $H=(H^{0},H^{-})^{T}$, $\Phi=(\Phi^{0},\Phi^{-})^{T}$ are all doublets under $SU(2)_{L}$, while $N_{R},N_{L},\chi_{R},\phi_{1},\phi_{2}$ are all singlets under the SM group. The particles in the dark sector have all non-vanishing $D$ numbers, while all of the SM particles have no $D$ numbers. Note that $N_{R}$ is filled into the SM sector but $N_{L}$ belongs to the dark sector, $N_{R}$ and $N_{L}$ will combine to form a heavy neutral Dirac fermion after the model symmetry breakings. $\chi_{R}$ is a neutral Majorana fermion, it will become the CDM. The dark neutral scalars $\phi_{1}$ and $\phi_{2}$ are applied to implement the spontaneous breakings of $U(1)_{D}$ and $U(1)_{B-L}$, while the dark doublet scalar $\Phi$ plays a key role in the generations of the neutrino mass and the matter-antimatter asymmetry. Finally, the model has also a hidden $Z_{2}$ symmetry, it is defined by the following discrete transform,
\ba
 f_{L}\rightarrow-f_{L},\hspace{0.3cm} f_{R}\rightarrow f_{R},\hspace{0.3cm} H\rightarrow-H,\hspace{0.3cm} \phi_{1}\rightarrow-\phi_{1},\hspace{0.3cm} \phi_{2}\rightarrow\phi_{2},\hspace{0.3cm} \Phi\rightarrow\Phi,
\ea
and all the gauge fields remain unchanged, where $f_{L,R}$ denote the left-handed and right-handed fermions in Tab. 1. The last line of Tab. 1 lists the $Z_{2}$ parity of each field. Note that $N_{L}$ and $\chi_{R}$ have the same gauge quantum numbers but they have opposite $Z_{2}$ parities. By virtue of the assignment of Tab. 1, it is easily verified that all of the chiral anomalies are completely cancelled in the model, namely the model is anomaly-free.

  Under the above symmetries, the invariant Lagrangian of the model is composed of the three parts of the gauge kinetic energy terms, the Yukawa couplings and the scalar potentials. The gauge kinetic energy terms are
\begin{alignat}{1}
 \mathscr{L}_{G}&=\mathscr{L}_{pure\:gauge}+\sum\limits_{f}i\,\overline{f}\,\gamma^{\mu}D_{\mu}f+\sum\limits_{S}(D^{\mu}S)^{\dagger}D_{\mu}S, \nonumber\\
 D_{\mu}&=\partial_{\mu}+i\left(g_{2}W_{\mu}^{i}\frac{\tau^{i}}{2}+g_{1}B_{\mu}\frac{Y}{2}+g_{0}Z'_{\mu}\frac{D}{2}\right),
\end{alignat}
where $f$ and $S$ respectively denote all kinds of the fermions and scalars in Tab. 1. $g_{0}$ and $Z'_{\mu}$ are the gauge coupling coefficient and gauge field which are associated with the local $U(1)_{D}$ symmetry. $\tau^{i}$ is the Paul matrices and the other notations are self-explanatory.

  The Yukawa couplings are
\begin{alignat}{1}
 \mathscr{L}_{Y}=&\:\overline{l_{L}}Y_{e}e_{R}\,i\tau_{2}H^{*}+\overline{l_{L}}Y_{1}N_{R}H+\overline{l_{L}}Y_{2}C\overline{N_{L}}^{T}\Phi \nonumber\\
 &+\overline{N_{L}}Y_{N}N_{R}\,\phi_{1}+\frac{1}{2}\,\overline{N_{L}}Y'_{N}C\overline{N_{L}}^{T}\phi_{2}+\frac{1}{2}\,\chi_{R}^{T}CY_{\chi}\chi_{R}\,\phi^{*}_{2}+h.c.\,,
\end{alignat}
where $C$ is the charge conjugation matrix and $C\overline{N_{L}}^{T}=N^{c}_{R}, \chi_{R}^{T}C=\overline{\chi^{c}_{L}}$. Note that the $Z_{2}$ symmetry of Eq. (1) forbids the explicit mass term $\overline{N_{L}}M\chi_{R}$ even though it satisfies all the gauge symmetries. This will guarantee the  stability of $\chi_{R}$ since it can not mix with the other fermions. These coupling parameters in Eq. (3), $Y_{e}, Y_{1}$, etc., are all $3\times3$ complex matrices in the flavour space, the leading elements of them take such moderate values as $[Y_{e},Y_{1},Y_{2}]\sim0.01$ and $[Y_{N},Y'_{N},Y_{\chi}]\sim0.1$. In addition, we can however choose such flavour basis in which $Y_{e},Y_{N},Y_{\chi}$ are simultaneously diagonal matrices, namely the mass eigenstate basis (see the following Eq. (14)), thus $Y_{1}$ and $Y_{2}$ certainly contain some irremovable complex phases, they eventually become $CP$-violating sources in the lepton sector in comparison with one in the quark sector. Eq. (3) will give rise to all kinds of the fermion masses after the scalar fields developing their non-vanishing vacuum expectation values. In particular, the spontaneous breakings of $U(1)_{D}$ and $U(1)_{B-L}$ will bring about significant results, the $Y_{1}$ and $Y_{2}$ terms can jointly lead to the tiny neutrino mass and the successful leptogenesis, while the $Y_{\chi}$ term will generate the CDM and yield correct annihilation cross-section.

  The full scalar potentials are
\begin{alignat}{1}
 V_{S}=&\:\mu^{2}_{\phi_{1}}\phi_{1}^{*}\phi_{1}+\mu^{2}_{\phi_{2}}\phi_{2}^{*}\phi_{2}+\mu^{2}_{H}H^{\dagger}H+\mu^{2}_{\Phi}\Phi^{\dagger}\Phi \nonumber\\
 &+\lambda_{\phi_{1}}(\phi_{1}^{*}\phi_{1})^{2}+\lambda_{\phi_{2}}(\phi_{2}^{*}\phi_{2})^{2}+\lambda_{H}(H^{\dagger}H)^{2}+\lambda_{\Phi}(\Phi^{\dagger}\Phi)^{2} \nonumber\\
 &+2\lambda_{0}\phi^{*}_{1}\phi_{1}\phi^{*}_{2}\phi_{2}+2(\lambda_{1}\phi^{*}_{1}\phi_{1}+\lambda_{2}\phi^{*}_{2}\phi_{2})H^{\dagger}H+2(\lambda_{3}\phi^{*}_{1}\phi_{1}+\lambda_{4}\phi^{*}_{2}\phi_{2})\Phi^{\dagger}\Phi \nonumber\\
 &+2\lambda_{5}H^{\dagger}H\Phi^{\dagger}\Phi +2\lambda_{6}H^{\dagger}\Phi\Phi^{\dagger}H-2\lambda_{7}(\phi_{1}\phi_{2}^{*}H^{\dagger}\Phi+h.c.).
\end{alignat}
All kinds of the parameters in Eq. (4) are chosen and required by the following conditions,
\begin{alignat}{1}
 &\mu_{\phi_{1}}\sim 10^{3}\;\mathrm{TeV}\;\mbox{and}\;\mu^{2}_{\phi_{1}}<0,\hspace{0.3cm} [\mu_{\phi_{2}},\mu_{H},\mu_{\Phi}]\sim 1\;\mathrm{TeV}\;\mbox{and}\;\mu^{2}_{\Phi}>0, \nonumber\\
 &[\lambda_{\phi_{1}},\lambda_{\phi_{2}},\lambda_{H},\lambda_{\Phi}]\sim 0.1\;\mbox{and they are all positive}, \nonumber\\
 &[\lambda_{0},\lambda_{1},\lambda_{2},\cdots,\lambda_{7}]\lesssim 10^{-6}\;\mbox{and $\lambda_{7}\sim 10^{-7}$ is positive}.
\end{alignat}
Here we assume that there is no $CP$ violation in the scalar sector, in addition, in Eq. (4) the self-interaction of each scalar is stronger but the interactions among them are feeble, so in Eq. (5) those self-coupling parameters are far larger than those interactive coupling parameters. $\mu^{2}_{\phi_{1}}<0$ and $\vert\mu^{2}_{\phi_{1}}\vert\gg[\mu^{2}_{\phi_{2}},\mu^{2}_{H},\mu^{2}_{\Phi}]$ will lead that $\phi_{1}$ first develops a non-vanishing vacuum expectation value at high scale, and then the other scalars are induced to develop non-vanishing vacuum expectation values at low scale via their coupling to $\phi_{1}$, eventually the symmetry breakings proceed along the chain of the following Eq. (9). $\mu^{2}_{\Phi}>0$ means that $\mu_{\Phi}$ is actually the original mass of the dark doublet scalar $\Phi$. In short, the conditions of Eq. (5) are natural and reasonable, they can sufficiently guarantee the vacuum stability and the spontaneous breakings of the model symmetries in the proper order.

  The vacua of the spontaneous breakings are along the directions of the neutral component of each scalar field. We can rigorously solve the $V_{S}$ minimum for Eq. (4), and then derive the vacuum configurations as follows,
\begin{alignat}{1}
 &\phi_{1}\rightarrow\frac{\phi_{1}^{0}+v_{1}+iG'^{0}}{\sqrt{2}}\,,\hspace{0.5cm} \phi_{2}\rightarrow\frac{\phi_{2}^{0}+v_{2}+iG^{0}}{\sqrt{2}}\,, \nonumber\\
 &H\rightarrow\left(\begin{array}{c}\frac{h^{0}+v_{H}+iG''^{0}}{\sqrt{2}}\\H^{-}\end{array}\right),\hspace{0.5cm} \Phi\rightarrow\left(\begin{array}{c}\frac{\Phi^{0}_{R}+v_{\Phi}+i\Phi^{0}_{I}}{\sqrt{2}}\\\Phi^{-}\end{array}\right), \nonumber\\
 &\langle\phi_{1}\rangle=\frac{v_{1}}{\sqrt{2}}\,,\hspace{0.3cm} \langle\phi_{2}\rangle=\frac{v_{2}}{\sqrt{2}}\,,\hspace{0.3cm} \langle H\rangle=\frac{v_{H}}{\sqrt{2}}\left(\begin{array}{c}1\\0\end{array}\right),\hspace{0.3cm} \langle\Phi\rangle=\frac{v_{\Phi}}{\sqrt{2}}\left(\begin{array}{c}1\\0\end{array}\right),
\end{alignat}
where the four vacuum expectation values are determined by the tadpole equations as follows,
\begin{alignat}{1}
 &\mu^{2}_{\phi_{1}}=-\left(\lambda_{\phi_{1}}v_{1}^{2}+\lambda_{0}v_{2}^{2}+\lambda_{1}v_{H}^{2}+\lambda_{3}v_{\Phi}^{2}\right)+\lambda_{7}\frac{v_{1}v_{2}v_{H}v_{\Phi}}{v_{1}^{2}}\,, \nonumber\\
 &\mu^{2}_{\phi_{2}}=-\left(\lambda_{0}v_{1}^{2}+\lambda_{\phi_{2}}v_{2}^{2}+\lambda_{2}v_{H}^{2}+\lambda_{4}v_{\Phi}^{2}\right)+\lambda_{7}\frac{v_{1}v_{2}v_{H}v_{\Phi}}{v_{2}^{2}}\,, \nonumber\\
 &\mu^{2}_{H}=-\left(\lambda_{1}v_{1}^{2}+\lambda_{2}v_{2}^{2}+\lambda_{H}v_{H}^{2}+(\lambda_{5}+\lambda_{6})v_{\Phi}^{2}\right)+\lambda_{7}\frac{v_{1}v_{2}v_{H}v_{\Phi}}{v_{H}^{2}}\,, \nonumber\\
 &\mu^{2}_{\Phi}=-\left(\lambda_{3}v_{1}^{2}+\lambda_{4}v_{2}^{2}+(\lambda_{5}+\lambda_{6})v_{H}^{2}+\lambda_{\Phi}v_{\Phi}^{2}\right)+\lambda_{7}\frac{v_{1}v_{2}v_{H}v_{\Phi}}{v_{\Phi}^{2}}\,.
\end{alignat}
A physics solution of the vacuum expectation values which fulfil the conditions of Eq (5) is a hierarchy such as
\ba
 v_{\Phi}\sim 0.5\:\mathrm{MeV}\ll v_{H}\sim v_{2}\sim 250\:\mathrm{GeV}\ll v_{1}\sim 2000\:\mathrm{TeV}.
\ea
In fact, $v_{H}=246$ GeV has been fixed by the electroweak physics, $v_{2}$ will be determined by the CDM physics, $v_{1}$ and $v_{\Phi}$ will jointly be determined by the tiny neutrino mass and the successful leptogenesis.

  According to the assignment of Tab. 1 and the relations of Eq. (8), the model symmetries are spontaneously broken step by step through the breaking chain as follows,
\begin{alignat}{1}
 &U(1)_{B-L}^{global}\otimes U(1)_{D}^{local}\otimes Z_{2}=U(1)_{(B-L)-D}^{global}\otimes U(1)_{(B-L)+D}^{local}\otimes Z_{2} \nonumber\\
 &\xrightarrow{\langle\phi_{1}\rangle\sim 10^{6}\mathrm{GeV}} U(1)_{(B-L)-D}^{global}\xrightarrow{\langle\phi_{2}\rangle\sim 10^{2}\mathrm{GeV}}nothing, \nonumber\\
 &SU(2)_{L}\otimes U(1)_{Y} \xrightarrow{\langle H\rangle\sim 10^{2}\mathrm{GeV}} U(1)_{em}\,.
\end{alignat}
At the first step, $\langle\phi_{1}\rangle\sim 10^{6}$ GeV breaks the local $U(1)_{(B-L)+D}$ and the discrete $Z_{2}$ but the global $U(1)_{(B-L)-D}$ is kept as a residual symmetry. The $\phi_{1}^{0}$ component becomes a massive real scalar boson around the $v_{1}$ scale, while the pseudo-scalar Goldstone boson $G'^{0}$ is eaten by the massless $Z'_{\mu}$, the latter becomes a massive gauge boson through the Higgs mechanism. In addition, the $Y_{N}$ term in Eq. (3) generates a heavy Dirac fermion mass around the $v_{1}$ scale by a combination of the neutral $N_{L}$ and $N_{R}$. At the second step, $\langle\phi_{2}\rangle\sim 10^{2}$ GeV violates the global $U(1)_{(B-L)-D}$, $\phi^{0}_{2}$ and $G^{0}$ in Eq. (6) respectively become a massive real scalar and a massless Goldstone boson, in addition, $\chi_{R}$ obtains a Majorana mass around the $v_{2}$ scale, it will become the CDM because of its characteristics (which will be discussed at IV Section). Note that the $Y'_{N}$ term in Eq. (3) also generates a Majorana mass of $N_{L}$ around the $v_{2}$ scale, but it is far smaller than the Dirac mass of $N$, so this can not change the nature of $N$ as a Dirac fermion. At the third step, the electroweak breaking is accomplished by $\langle H\rangle\sim 10^{2}$ GeV, the SM particles obtain their masses around the electroweak scale. It should be stressed that the $B-L-D$ violation occurs at the time close to the electroweak breaking due to $v_{2}\sim v_{H}$. Lastly, the above three breakings can also induce the neutral component of the dark doublet $\Phi$ developing a smaller $\langle\Phi\rangle\sim 0.5$ MeV via the $\lambda_{7}$ feeble coupling term in Eq (4), of course, $v_{\Phi}$ is too small to make an effect on the nature of the heavy $\Phi$ whose mass is $\sim\mu_{\Phi}$.

  After the above symmetry breakings are completed, these components of $G'^{0},G''^{0},H^{-}$ in Eq. (6) have been transformed into the gauge sector to generate $M_{Z'},M_{Z},M_{W}$ through the Higgs mechanism, therefore the scalar sector now includes four $CP$-even neutral bosons $\phi_{1}^{0},\phi_{2}^{0},h^{0},\Phi^{0}_{R}$, two $CP$-odd ones $G^{0},\Phi^{0}_{I}$, and a pair of charged bosons $\Phi^{\mp}$. The squared mass matrix of the $CP$-even $(\phi_{1}^{0},\phi_{2}^{0},h^{0},\Phi^{0}_{R})$ and one of the $CP$-odd $(G^{0},\Phi^{0}_{I})$, and the squared mass of the charged $\Phi^{\mp}$, are together given as follows,
\begin{alignat}{1}
&M_{+}^{2}=\left(\begin{array}{cccc}2\lambda_{\phi_{1}}v_{1}^{2}+\frac{\Omega}{v_{1}^{2}}
 &2\lambda_{0}v_{1}v_{2}-\lambda_{7}v_{H}v_{\Phi}&2\lambda_{1}v_{1}v_{H}-\lambda_{7}v_{2}v_{\Phi}
 &2\lambda_{3}v_{1}v_{\Phi}-\lambda_{7}v_{2}v_{H}\\
 \ldots&2\lambda_{\phi_{2}}v_{2}^{2}+\frac{\Omega}{v_{2}^{2}}&2\lambda_{2}v_{2}v_{H}-\lambda_{7}v_{1}
 v_{\Phi}&2\lambda_{4}v_{2}v_{\Phi}-\lambda_{7}v_{1}v_{H}\\
 \ldots&\ldots&2\lambda_{H}v_{H}^{2}+\frac{\Omega}{v_{H}^{2}}&2(\lambda_{5}+\lambda_{6})v_{H}
 v_{\Phi}-\lambda_{7}v_{1}v_{2}\\
 \ldots&\ldots&\ldots&2\lambda_{\Phi}v_{\Phi}^{2}+\frac{\Omega}{v_{\Phi}^{2}}\end{array}\right), \nonumber\\
&M_{-}^{2}=\left(\begin{array}{cc}\frac{\Omega}{v_{2}^{2}}&-\lambda_{7}v_{1}v_{H}\\
 -\lambda_{7}v_{1}v_{H}&\frac{\Omega}{v_{\Phi}^{2}}\end{array}\right)
 \xrightarrow{diagonalizing}\left(\begin{array}{cc}0&0\\0&\frac{\Omega}{v_{2}^{2}}+\frac{\Omega}{v_{\Phi}^{2}}\end{array}\right), \nonumber\\
&M_{\Phi^{\mp}}^{2}=-\lambda_{6}v_{H}^{2}+\frac{\Omega}{v_{\Phi}^{2}}\,,
\end{alignat}
where $\Omega=\lambda_{7}v_{1}v_{2}v_{H}v_{\Phi}$. $M_{+}^{2}$ is obviously an approximately diagonal matrix because those non-diagonal elements are far smaller than those diagonal elements, of course, this arises from only feeble coupling among the scalar fields, so we can safely neglect the mixing among the $CP$-even bosons, for example, the mixing angle between $\phi^{0}_{2}$ and $h^{0}$ is $\sim\frac{\lambda_{2}v_{2}v_{H}}{\lambda_{\phi_{2}}v_{2}^{2}-\lambda_{H}v_{H}^{2}}\ll 1$ due to $\lambda_{2}\ll 1$. In $M_{-}^{2}$, the mixing angle between $G^{0}$ and $\Phi^{0}_{I}$ is $\sim\frac{v_{\Phi}}{v_{2}}\ll1$, however, $G^{0}$ indeed becomes a zero mass Goldstone boson because the determinant of $M_{-}^{2}$ is vanishing, this is of course an inevitable outcome of the global $B-L-D$ breaking. Eq. (10) indicates that $\Phi^{\mp},\Phi^{0}_{I},\Phi^{0}_{R}$ have nearly the same squared mass $\frac{\Omega}{v^{2}_{\Phi}}=\frac{\lambda_{7}v_{1}v_{2}v_{H}}{v_{\Phi}}$ on account of Eq. (5) and Eq. (8), therefore the neutral and charged components of the dark doublet $\Phi$ actually keep a degenerating mass in despite of their tiny splits.

  In the gauge sector, the squared mass matrix of the three neutral gauge bosons $(B_{\mu},W^{3}_{\mu},Z'_{\mu})$ and the squared mass of the charged gauge bosons $W_{\mu}^{\pm}$ are given by
\begin{alignat}{1}
 M_{NGB}^{2}=&\frac{1}{4}\left(\begin{array}{ccc}g_{1}^{2}(v_{H}^{2}+v_{\Phi}^{2})&-g_{1}g_{2}(v_{H}^{2}+v_{\Phi}^{2})&-g_{1}g_{0}v_{\Phi}^{2}\\
 -g_{1}g_{2}(v_{H}^{2}+v_{\Phi}^{2})&g_{2}^{2}(v_{H}^{2}+v_{\Phi}^{2})&g_{2}g_{0}v_{\Phi}^{2}\\
 -g_{1}g_{0}v_{\Phi}^{2}&g_{2}g_{0}v_{\Phi}^{2}&g_{0}^{2}(v_{1}^{2}+4v_{2}^{2})\end{array}\right)\nonumber\\
 \xrightarrow{diagonalizing}&\left(\begin{array}{ccc}M^{2}_{A_{\mu}}=0&0&0\\
 0&M^{2}_{Z_{\mu}}=\frac{g_{2}^{2}+g_{1}^{2}}{4}(v_{H}^{2}+v_{\Phi}^{2})&0\\
 0&0&M^{2}_{Z'_{\mu}}=\frac{g_{0}^{2}}{4}(v_{1}^{2}+4v_{2}^{2})\end{array}\right), \nonumber\\
 M^{2}_{W_{\mu}}=&\frac{g_{2}^{2}}{4}(v_{H}^{2}+v_{\Phi}^{2}),
\end{alignat}
where the photon field $A_{\mu}$ is massless because the determinant of $M_{NGB}^{2}$ is vanishing. The mixing angle between $Z'_{\mu}$ and $B_{\mu}$ is $\sim\frac{g_{1}v_{\Phi}^{2}}{g_{0}v_{1}^{2}}$, the mixing angle between $Z'_{\mu}$ and $W_{\mu}^{3}$ is $\sim\frac{g_{2}v_{\Phi}^{2}}{g_{0}v_{1}^{2}}$, obviously, they are nearly zero, so we can leave them out. The mixing angle between $B_{\mu}$ and $W_{\mu}^{3}$ is $\cos\theta_{w}=\frac{g_{2}}{\sqrt{g_{2}^{2}+g_{1}^{2}}}$, which is namely the weak mixing angle of the SM. In addition, $M_{W}=\frac{g_{2}v_{H}}{2}(1+\frac{v_{\Phi}^{2}}{v_{H}^{2}})^{\frac{1}{2}}$ and $M_{Z}=\frac{M_{W}}{\cos\theta_{w}}$ have only a very tiny correction $\sim\frac{v_{\Phi}^{2}}{v_{H}^{2}}$ to the SM values, which is very difficult to be detected.

  In the fermion sector, the mass matrix of the neutral fermions is written as
\begin{alignat}{1}
&\frac{1}{2}\,(\overline{\nu_{L}},\overline{N_{L}},N_{R}^{T}C,\chi_{R}^{T}C)\left(\begin{array}{cccc}0&\frac{v_{\Phi}Y_{2}}{\sqrt{2}}&\frac{v_{H}Y_{1}}{\sqrt{2}}&0\\\frac{v_{\Phi}Y_{2}^{T}}{\sqrt{2}}&\frac{v_{2}Y'_{N}}{\sqrt{2}}&\frac{v_{1}Y_{N}}{\sqrt{2}}&0\\\frac{v_{H}Y_{1}^{T}}{\sqrt{2}}&\frac{v_{1}Y_{N}^{T}}{\sqrt{2}}&0&0\\0&0&0&\frac{v_{2}Y_{\chi}}{\sqrt{2}}\end{array}\right)\left(\begin{array}{c}C\overline{\nu_{L}}^{T}\\C\overline{N_{L}}^{T}\\N_{R}\\\chi_{R}\end{array}\right) \nonumber\\
\xrightarrow{diagonalizing}&-\frac{1}{2}\,(\overline{\nu'_{L}},\overline{N'_{L}},N'^{T}_{R}C,\chi_{R}^{T}C)\left(\begin{array}{cccc}M_{\nu}&0&0&0\\0&0&M_{N}&0\\0&M_{N}^{T}&0&0\\0&0&0&M_{\chi}\end{array}\right)\left(\begin{array}{c}C\overline{\nu'_{L}}^{T}\\C\overline{N'_{L}}^{T}\\N'_{R}\\\chi_{R}\end{array}\right) \nonumber\\
&=-\frac{1}{2}\,\overline{\nu'_{L}}M_{\nu}C\overline{\nu'_{L}}^{T}-\overline{N'_{L}}M_{N}N'_{R}-\frac{1}{2}\,\chi_{R}^{T}CM_{\chi}\chi_{R}\,,
\end{alignat}
where $M_{\nu},M_{N},M_{\chi}$ are given by Eq. (14) below, and $\nu',N'$ are new fermion fields after the flavour rotation. Because of $v_{2}Y'_{N}\ll v_{1}Y_{N}$, $N_{L}$ and $N_{R}$ actually combine into a heavy Dirac fermion. In addition, $\chi_{R}$ has no mixing with the other neutral fermions due to the $Z_{2}$ symmetry in Eq. (1), this leads that it eventually becomes the CDM. At the low energy, the heavy Dirac fermion $N$ has decoupled and it can be integrated out from the two terms of $Y_{1}$ and $Y_{2}$ in Eq. (3), thus we can obtain an effective coupling of the left-handed doublet lepton
\ba
 \mathscr{L}_{eff}=\frac{1}{2}\,\overline{l_{L}}(HY_{1}M_{N}^{-1}Y_{2}^{T}\Phi^{T}+\Phi Y_{2}M_{N}^{-1T}Y_{1}^{T}H^{T})C\overline{l_{L}}^{T}+h.c.\,.
\ea
According to the assignment of Tab. 1, this effective coupling explicitly violates both one unit of $B-L$ number and one unit of $D$ number, namely violates two unit of $B-L+D$ number but conserves $B-L-D$ number. In fact, Eq. (13) is exactly a common origin of both the tiny neutrino mass and the leptogenesis.

  We can now give all of the model particle masses as follows,
\begin{alignat}{1}
 &M_{Z'_{\mu}}=\frac{v_{1}g_{0}}{2}\,,\hspace{0.3cm} M_{\phi^{0}_{1}}=v_{1}\sqrt{2\lambda_{\phi_{1}}}\,,\hspace{0.3cm} M_{\phi^{0}_{2}}=v_{2}\sqrt{2\lambda_{\phi_{2}}}\,,\hspace{0.3cm} M_{G^{0}}=0, \nonumber\\
 &M_{h^{0}}=v_{H}\sqrt{2\lambda_{H}}\,,\hspace{0.5cm} M_{\Phi}=\sqrt{\frac{\lambda_{7}v_{1}v_{2}v_{H}}{v_{\Phi}}}\approx\mu_{\Phi}\sqrt{1+\frac{\lambda_{3}v_{1}^{2}}{\mu_{\Phi}^{2}}}\,, \nonumber\\
 &M_{N}=-\frac{v_{1}}{\sqrt{2}}Y_{N},\hspace{0.5cm} M_{\chi}=-\frac{v_{2}}{\sqrt{2}}Y_{\chi},\hspace{0.5cm} M_{e}=\frac{v_{H}}{\sqrt{2}}Y_{e}, \nonumber\\
 &M_{\nu}=-\frac{v_{H}v_{\Phi}}{2}(Y_{1}M_{N}^{-1}Y_{2}^{T}+Y_{2}M_{N}^{-1T}Y_{1}^{T})=\frac{v_{H}^{2}v_{2}\lambda_{7}}{\sqrt{2}M_{\Phi}^{2}}(Y_{1}Y_{N}^{-1}Y_{2}^{T}+Y_{2}Y_{N}^{-1T}Y_{1}^{T}).
\end{alignat}
$M_{h^{0}}$ is namely the Higgs boson mass of the SM, which has been measured as $M_{h^{0}}\approx125$ GeV \cite{1}. $M_{\Phi}$ is about several TeVs, which is close to its original mass $\mu_{\Phi}$. The tiny Majorana mass of the SM neutrino is generated by Eq. (12) or equivalently by Eq. (13), it is jointly suppressed by both the small $v_{\Phi}$ and the heavy $M_{N}$. The second equality of $M_{\nu}$ is obtained by use of the $M_{N}$ equality and the $M_{\Phi}$ one, from a further point of view, the tiny $M_{\nu}$ essentially arises from the $\lambda_{7}$ feeble coupling. Therefore this is a hybrid see-saw mechanism, in particular, it is realized at the TeV scale and it is possibly tested in the future by measuring the parameters in the second equality of $M_{\nu}$. Finally, the neutrino mass matrix $M_{\nu}$ bears full information of the neutrino mass and the lepton mixing.

  Based on Eq. (5), Eq. (8) and Eq. (14), in addition, the mass hierarchy of $N_{1,2,3}$ and one of $\chi_{1,2,3}$ are all taken into account, thus we can infer that the mass spectrum of the model particles are such relations as (GeV as unit),
\begin{alignat}{1}
 &M_{G^{0}}<M_{\nu}\sim 10^{-10}\ll M_{e}<M_{\chi_{1}}\sim 10<M_{\chi_{2}}<M_{\chi_{3}}\sim M_{h^{0}}\sim M_{\phi^{0}_{2}}\sim 10^{2} \nonumber\\
 &<M_{\Phi}\sim 10^{3}<M_{N_{1}}\sim 10^{5}<M_{N_{2}}<M_{N_{3}}\sim M_{\phi^{0}_{1}}\sim M_{Z'_{\mu}}\sim 10^{6}.
\end{alignat}
This is easily fulfilled by choosing some suitable values of the coupling parameters in Eq. (14). However, the mass relations of Eq. (15) will successfully lead to the leptogenesis and the CDM. Finally, it is worth emphasizing that there are not any super-high scale physics in the model.

\vspace{0.6cm}
\noindent\textbf{III. Leptogenesis}

\vspace{0.3cm}
  The model can account for the baryon asymmetry through the leptogenesis. Below the $v_{1}$ scale but above the $v_{2}$ scale, $B-L+D$ is violated by $\langle\phi_{1}\rangle$, but $B-L-D$ is conserved, moreover, is anomaly-free. The dark doublet $\Phi$ at the TeV scale has two decay modes on the basis of the model couplings and the (15) relations, (i) the two-body decay $\Phi\rightarrow H+\phi_{2}$ via the $\lambda_{7}$ term in Eq. (4), (ii) the three-body decay $\Phi\rightarrow l_{\alpha}+l_{\beta}+\overline{H}$ via Eq. (13). Fig. 1 shows the tree and loop diagrams of $\Phi\rightarrow l_{\alpha}+l_{\beta}+\overline{H}$.
\begin{figure}
 \centering
 \includegraphics[totalheight=7cm]{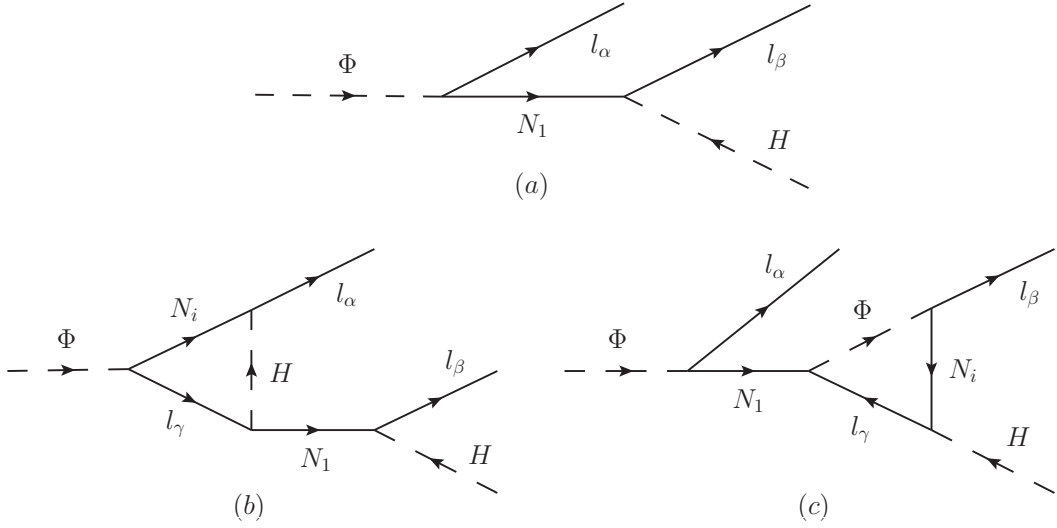}
 \caption{The tree and loop diagrams of the decay $\Phi\rightarrow l_{\alpha}+l_{\beta}+\overline{H}$ violating $B-L+D$ but conserving $B-L-D$, this decay has the $CP$ asymmetry and is out-of-equilibrium, which eventually leads to the matter-antimatter asymmetry.}
\end{figure}
Note that the three-body decay is mainly mediated via $N_{1}$ as shown Fig. 1, the diagrams in which the decay is mediated via $N_{2,3}$ can be neglected because they are greatly suppressed due to $M^{-4}_{N_{2,3}}\ll M^{-4}_{N_{1}}$. In term of the quantum number assignment in Tab. 1, explicitly, this process of Fig. 1 simultaneously violates ``$-1$" unit of $B-L$ number and ``$-1$" unit of $D$ number, namely $\triangle(B-L)=-1$ and $\triangle D=-1$, so one can obtain $\triangle(B-L+D)=-2$ and $\triangle(B-L-D)=0$ as expected. On the other hand, the decay of Fig. 1 has a $CP$ asymmetry and is out-of-equilibrium, but the decay of $\Phi\rightarrow H+\phi_{2}$ has no these features. However, the two-body decay rate is much larger than the three-body one, therefore the total decay width of $\Phi$ is approximately equal to the decay width of $\Phi\rightarrow H+\phi_{2}$.

  Because the two coupling matrices of $Y_{1}$ and $Y_{2}$ contain the $CP$-violating sources, the decay rate of $\Phi\rightarrow l_{\alpha}+l_{\beta}+\overline{H}$ is different from one of its $CP$-conjugate process $\overline{\Phi}\rightarrow\overline{l_{\alpha}}+\overline{l_{\beta}}+H$ through the interference between the tree diagram and the loop one. The $CP$ asymmetry of the two decay rates is defined and calculated as follows,
\begin{alignat}{1}
 &\varepsilon=\frac{\Gamma^{+}-\Gamma^{-}}{\Gamma_{\Phi}^{total}}=\frac{(Y_{1}^{\dagger}Y_{1})_{11}M^{4}_{\Phi}\sum\limits_{i\neq1}M_{N_{i}}Im[(Y_{1}^{\dagger}Y_{2})_{1i}(Y_{2}^{\dagger}Y_{1})_{1i}]}{768\pi^{3}M^{3}_{N_{1}}(\lambda_{7}v_{1})^{2}}\,, \nonumber\\
 &\Gamma^{\pm}=\sum\limits_{\alpha,\beta}\Gamma(\frac{\Phi\rightarrow l_{\alpha}+l_{\beta}+\overline{H}}{\overline{\Phi}\rightarrow\overline{l_{\alpha}}+\overline{l_{\beta}}+H})=\Gamma_{tree}+\Gamma_{loop}^{\pm}\,,\hspace{0.5cm} \Gamma_{tree}=\frac{(Y_{1}^{\dagger}Y_{1})_{11}(Y_{2}^{\dagger}Y_{2})_{11}M^{3}_{\Phi}}{1536\pi^{3}M^{2}_{N_{1}}}\,, \nonumber\\
 &\Gamma_{\Phi}^{total}\approx\Gamma(\Phi\rightarrow H+\phi_{2})=\frac{(\lambda_{7}v_{1})^{2}}{8\pi M_{\Phi}}\,.
\end{alignat}
A careful calculation shows that the imaginary part of the loop integration factor of the (b) diagram is derived from the three-point function $Im[(C_{0}+C_{12})(M^{2}_{l_{\alpha}},s_{12},M^{2}_{\Phi},M^{2}_{N_{i}},M^{2}_{H},M^{2}_{l_{\gamma}})]$ $=\frac{2\pi i}{M^{2}_{\Phi}-s_{12}}$, where $s_{12}=(p_{\overline{H}}+p_{l_{\beta}})^{2}$, but the (c) diagram has in fact no contribution to $\varepsilon$ because the imaginary part of its three-point function is vanishing. According to the discussions in Section II, there are $Y_{1}\sim Y_{2}\sim0.01$ and $\lambda_{7}\sim10^{-7}$, then we can roughly estimate $\varepsilon\sim 10^{-8}$ by use of Eq. (8) and Eq. (15), it is exactly a reasonable and suitable value for the leptogenesis.

  A further calculation shows that the decay rate $\Gamma^{\pm}$ in Eq. (16) is smaller than the  Hubble expansion rate of the universe, namely
\ba
 \Gamma^{\pm}\approx\Gamma_{tree}<H(M_{\Phi})=\frac{1.66\sqrt{g_{*}}M^{2}_{\Phi}}{M_{Pl}}\,,
\ea
where $M_{Pl}=1.22\times10^{19}$ GeV and $g_{*}$ is the effective number of relativistic degrees of freedom. Therefore the decay process of Fig. 1 is actually out-of-equilibrium. At the scale of $T=M_{\Phi}$, the relativistic states include all of the SM particles, and $\phi_{2}$ and $\chi_{i}$ in the dark sector, so one can figure out $g_{*}=114$ in Eq. (17). By this time, we have completely demonstrated that the decay process of Fig. 1 is able to satisfy Sakharov's three conditions \cite{16}.

  The above discussions are now integrated together, as a result, the decay of Fig. 1 certainly generates the following asymmetries \cite{17},
\begin{alignat}{1}
 &Y_{B-L}=\frac{n_{B-L}-\overline{n}_{B-L}}{s}=\kappa\frac{\triangle(B-L)\,\varepsilon}{g_{*}}=\kappa\frac{(-1)\varepsilon}{g_{*}}\,, \nonumber\\
 &Y_{D}=\frac{n_{D}-\overline{n}_{D}}{s}=\kappa\frac{\triangle D\,\varepsilon}{g_{*}}=\kappa\frac{(-1)\varepsilon}{g_{*}}\,, \nonumber\\
 &Y_{B-L+D}=\kappa\frac{\triangle(B-L+D)\,\varepsilon}{g_{*}}=\kappa\frac{(-2)\varepsilon}{g_{*}}\,, \nonumber\\
 &Y_{B-L-D}=\kappa\frac{\triangle(B-L-D)\,\varepsilon}{g_{*}}=0,
\end{alignat}
where $s$ is the entropy density and $\kappa$ is a dilution factor. The dilution is mostly from the  inverse decay. For the three-body inverse decay, the dilution effect is very weak if the departure from thermal equilibrium is severe, so we can take $\kappa\approx1$ in Eq. (18). Note that the dilution effect from $N_{1}\rightarrow\Phi+\overline{l_{\alpha}}$ is almost nothing because $N_{1}$ has early decoupled and its number density is exponentially suppressed by $\frac{M_{N_{1}}}{M_{\Phi}}\sim20$ compared to the $\Phi$ one.

  At the high energy, the SM sector and the dark sector are connected each other via the heavy particles $Z_{\mu}',N_{i},\phi_{1}^{0},\Phi$ mediating. As the universe temperature falls below $M_{\Phi}$, all of the heavy particles have completely decayed and decoupled, thus the connection between the SM sector and the dark sector is gradually discontinued, eventually, at the low energy the SM sector and the dark sector are isolated from each other, the surviving $\phi_{1}$ and $\chi_{i}$ in the dark sector constitute a dark world. As a consequence, the $Y_{B-L}$ asymmetry is totally deposited in the SM sector because $\phi_{1}$ and $\chi_{i}$ in the dark sector have vanishing $B-L$ numbers (see Tab. 1), while the $Y_{D}$ asymmetry is totally deposited in the dark sector because all of the SM particles have no $D$ numbers (see Tab. 1). In other words, at the low energy, the SM sector has the $B-L$ asymmetry without the $D$ asymmetry, while the dark sector has the $D$ asymmetry without the $B-L$ asymmetry. However, the total $Y_{B-L-D}$ asymmetry in the two sectors is kept to be zero.

  When the universe temperature drops to the energy scale $v_{2}\sim v_{H}\sim250$ GeV, the global $B-L-D$ symmetry is broken because $\phi_{2}$ developing $\langle\phi_{2}\rangle$ violates two unit of $D$ number. The original complex $\phi_{2}$ is now decomposed into the two neutral real $\phi_{2}^{0},G^{0}$ which are namely their own antiparticles, and also $\chi_{R}$ becomes a neutral Majorana fermion, namely $\chi=\chi^{c}$. Consequently, the $Y_{D}$ asymmetry between dark particles and dark antiparticles is totally erased, or rather it is automatically vanishing. Although the total $B-L-D$ conservation in the two sectors is violated, the SM sector always conserves the $B-L$ number, therefore the $Y_{B-L}$ asymmetry in the SM sector is unaffected and unchanged. At this temperature of $v_{2}\sim v_{H}\sim250$ GeV, obviously, the electroweak sphaleron process can fully put into effect in the SM sector \cite{18}, thus a part of the $Y_{B-L}$ asymmetry is converted into the baryon asymmetry. This is given by the following relation,
\ba
 \eta_{B}=\frac{n_{B}-\overline{n}_{B}}{n_{\gamma}}=7.04\,c_{s}Y_{B-L}\approx 6.2\times10^{-10},
\ea
where $c_{s}=\frac{28}{79}$ is the sphaleron conversion coefficient. $7.04$ is a ratio of the entropy density to the photon number density. $6.2\times10^{-10}$ is the current value of the baryon asymmetry \cite{19}. Note that $\chi_{i},\phi_{2}^{0},G^{0}$ in the dark sector do not at all participate in the sphaleron process since they are all singlets under the SM group and isolated from the SM sector. When the universe temperature falls below $T\sim100$ GeV, the sphaleron process is closed and the baryon asymmetry is kept up to the present day. Finally, it should be stressed that the leptogenesis is elegantly achieved at the TeV scale in the model.

\vspace{0.6cm}
\noindent\textbf{IV. Dark Matter}

\vspace{0.3cm}
  When the global symmetry of $U(1)_{B-L-D}$ is broken by $\langle\phi_{2}\rangle$ at the scale $v_{2}\sim v_{H}$, the dark particles $\chi_{i},\phi_{2}^{0},G^{0}$ are at first in thermal equilibrium in the dark world. At a later time $\phi_{2}^{0}$ can completely decay into a pair of $\chi_{i}$ or $G^{0}$. However $\chi_{i}$ is a stable particle because it is protected by the two factors. i) The model gauge symmetries prevent it from coupling to the other particles except $\phi_{2}$ (see Eq. (3)), so it can not decay. ii) The $Z_{2}$ symmetry in Eq. (1) forbids the explicit mass term $\overline{N_{L}}M\chi_{R}$, so it can not mix with the other fermions. Therefore $\chi_{i}$ is a stable WIMP. Among the three generation of $\chi_{1,2,3}$, a pair of heavier $\chi_{2,3}$ mainly annihilate into a pair of the lightest $\chi_{1}$ via the $G^{0}$ mediator, as shown (a) in Fig. 2. At the last step a pair of $\chi_{1}$ annihilate into a pair of $G^{0}$ by the two modes of (b) and (c) in Fig. 2.
\begin{figure}
 \centering
 \includegraphics[totalheight=8cm]{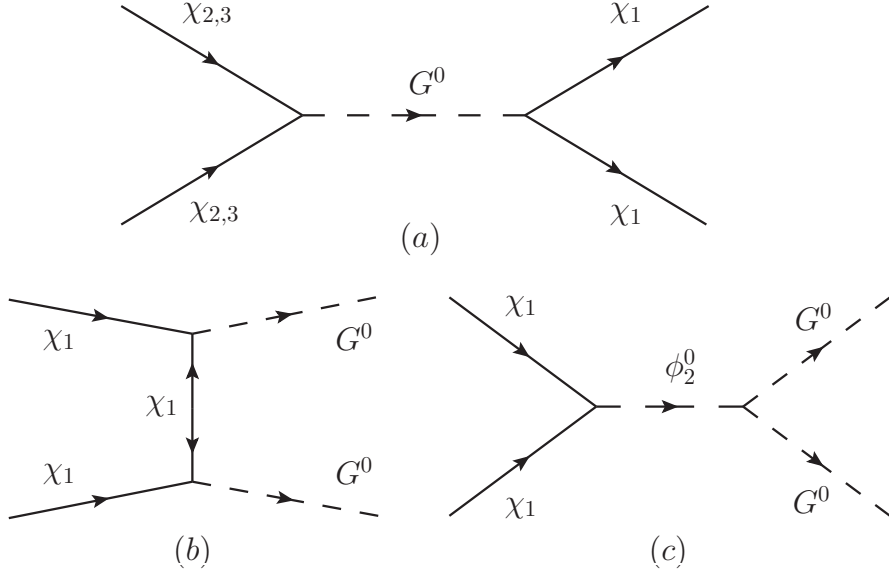}
 \caption{(a) A pair of heavier $\chi_{2,3}$ annihilating into a pair of the lightest $\chi_{1}$, (b) and (c) A pair of the CDM $\chi_{1}$ annihilating into a pair of Goldstone bosons $G^{0}$, which correctly leads to the ``WIMP Miracle".}
\end{figure}
A careful analysis shows that the annihilation cross-section of $\chi_{2,3}$ in (a) diagram is much larger than one of $\chi_{1}$ in (b) and (c) diagrams, therefore, the $\chi_{2,3}$ decoupling is much earlier than the $\chi_{1}$ one, and the relic abundance of $\chi_{2,3}$ is much smaller than one of $\chi_{1}$. Consequently, $\chi_{1}$ becomes the principal particle of the CDM, while $\chi_{2,3}$ only bears a tiny part of the CDM budget, in addition, $G^{0}$ becomes a dark background radiation. In short, $\chi_{1}$ is a desirable candidate of the CDM because its natures and relic abundance are very well consistent with ones of the CDM.

  $\chi_{1}$ becomes a non-relativistic particle when the temperature falls below $M_{\chi_{1}}$. It has two annihilation channels. (i) $\chi_{1}+\chi_{1}\rightarrow G^{0}+G^{0}$ via the $\chi_{1}$ t-channel mediation, as shown (b) in Fig. 2, note that the $\chi_{1}$ u-channel mediation is not shown in Fig. 2 but it is included in the following calculation. (ii) $\chi_{1}+\chi_{1}\rightarrow G^{0}+G^{0}$ via the $\phi_{2}^{0}$ s-channel mediation, as shown (c) in Fig. 2. The total annihilation rate of $\chi_{1}$ is calculated as follows,
\begin{alignat}{1}
 &\Gamma(\chi_{1}+\chi_{1}\rightarrow G^{0}+G^{0})=\langle\sigma v\rangle n_{\chi_{1}},\hspace{0.5cm} n_{\chi_{1}}=2\left(\frac{M_{\chi_{1}}T}{2\pi}\right)^{\frac{3}{2}}e^{-\frac{M_{\chi_{1}}}{T}}, \nonumber\\
 &\sigma v=\frac{M^{2}_{\chi_{1}}}{128\pi v_{2}^{4}}\left[\left(\frac{1}{(1-4r)^{2}}+\frac{3-4r}{3(1-4r)}\right)v^{2}-\frac{3-4r}{30(1-4r)}\,v^{4}+\cdots\right], \nonumber\\
 &\langle\sigma v\rangle=a+b\,\langle v^{2}\rangle+c\,\langle v^{4}\rangle+\cdots\approx a+b\frac{6\,T}{M_{\chi_{1}}}\,, \nonumber\\
 &v=2\sqrt{1-\frac{4M^{2}_{\chi_{1}}}{s}}\,,\hspace{0.5cm} r=\frac{M^{2}_{\chi_{1}}}{M^{2}_{\phi_{2}^{0}}}\,,
\end{alignat}
where $a$ and $b$ are determined by the $\sigma v$ equation, $v$ is a relative velocity of two annihilating particles. Eq. (20) clearly shows that the annihilation cross-section essentially arises from a p-wave contribution. In view of Eq. (8) and Eq. (15), the thermal average on the annihilation cross-section is actually $\langle\sigma v\rangle\sim10^{-9}$ $\mathrm{GeV}^{-2}$, which is exactly a weak interaction cross-section. This naturally reproduces the so-called ``WIMP Miracle" \cite{20}.

  As the universe temperature decreasing, the annihilation rate of $\chi_{1}$ becomes smaller than the Hubble expansion rate of the universe, then the annihilation is out-of equilibrium and $\chi_{1}$ is decoupling. The freeze-out temperature is determined by
\begin{alignat}{1}
 &\Gamma(T_{f})=H(T_{f})=\frac{1.66\sqrt{g_{*}(T_{f})}\,T_{f}^{2}}{M_{Pl}}\,, \nonumber\\
\Longrightarrow &x=\frac{T_{f}}{M_{\chi_{1}}}\approx\left(17.6+ln\frac{M_{\chi_{1}}}{\sqrt{g_{*}(T_{f})x}}+ln\frac{\langle\sigma v\rangle}{10^{-10}\:\mathrm{GeV^{-2}}}\right)^{-1}.
\end{alignat}
After $\chi_{1}$ is frozen out, its numbers in the comoving volume has no change any more. The current relic abundance of $\chi_{1}$ can be calculated by the following equation \cite{20},
\ba
 \Omega_{\chi_{1}}h^{2}=\frac{0.85\times10^{-10}\:\mathrm{GeV}^{-2}}{\sqrt{g_{*}(T_{f})}\,x(a+3bx)}\approx 0.12.
\ea
$0.12$ is the current abundance of the CDM \cite{21}. Obviously, both $M_{\chi_{1}}$ and $v_{2}$ are jointly in charge of the final results of Eq. (21) and Eq. (22). Provided $M_{\chi_{1}}\sim60$ GeV and $v_{2}\sim250$ GeV, then the solution of Eq. (21) is $x\sim\frac{1}{25}$, namely the freeze-out temperature is $T_{f}\sim2.3$ GeV. At this temperature the relativistic particles include $photon,gluon,\nu^{0},e^{-},\mu^{-},u,d,s$ and $G^{0}$, so we can figure out $g_{*}(T_{f})=62.75$. Finally, we can correctly reproduce $\Omega_{\chi_{1}}h^{2}\sim0.12$ by Eq. (20) and Eq. (22).

  Obviously, the $G^{0}$ decoupling is exactly at the same temperature as the CDM $\chi_{1}$ decoupling. Since $G^{0}$ is massless and a relativistic decoupling, nowadays it should become a dark background radiation, which is analogous to the CMB photon in the visible world. Because the $G^{0}$ decoupling is much earlier than the neutrino decoupling and the photon one, its effective temperature is lower than the neutrino effective temperature and the CMB photon temperature. As a result, the current abundance of $G^{0}$, $\Omega_{G^{0}}$, is smaller than the neutrino abundance $\Omega_{\nu}\approx1.7\times10^{-3}$ and the CMB photon abundance $\Omega_{\gamma}\approx5\times10^{-5}$, refer to the review of cosmological parameters in \cite{1}. However, $G^{0}$ is in the dark sector and does not interact with the SM matters, so we can not detect it through the ordinary methods.

  Two CDM $\chi_{1}$ can interact each other through the long range exchange of $G^{0}$, but its effective potential is a repulsive force since $\chi_{1}$ is its own antiparticle, therefore there are not any bound states for the CDM $\chi_{1}$, they can only happen elastic scattering. When its reaction rate is smaller than the universe expansion rate, this elastic scattering will be frozen out and closed. The frozen-out temperature is determined by
\ba
\Gamma(\chi_{1}+\chi_{1}\rightarrow \chi_{1}+\chi_{1})=\langle\sigma v\rangle n_{\chi_{1}}\approx\frac{M^{2}_{\chi_{1}}}{32\pi v_{2}^{4}}\,\overline{v}\,n_{\chi_{1}}=H(T),
\ea
where $\overline{v}\approx\sqrt{\frac{2\,T}{\pi M_{\chi_{1}}}}$ is an average relative velocity. By use of the parameter values in Eq. (24) below, we can calculate that the frozen-out temperature is $\frac{T}{M_{\chi_{1}}}\approx\frac{1}{24}$. It is approximately equal to the $\chi_{1}$ decoupling temperature $\frac{T_{f}}{M_{\chi_{1}}}\approx\frac{1}{25}$. The reason for this is obviously that both the scattering cross-section in Eq. (23) and the annihilation cross-section in Eq. (20) are a weak interaction cross-section. Therefore, the elastic scattering between the CDM $\chi_{1}$ is actually frozen out at the same time when they are decoupling. Thereafter the CDM $\chi_{1}$ are completely free particles except the gravitational influence. In conclusion, the model can simply account for the CDM, in particular, naturally explain the ``WIMP Miracle".

\vspace{0.6cm}
\noindent\textbf{V. Numerical Results and Discussions}

\vspace{0.3cm}
  We now show some concrete numerical results of the model. All of the SM parameters have been fixed by the current experimental data \cite{1}. Some new parameters in the model can be determined by the current data of the tiny neutrino mass, the baryon asymmetry, and the CDM abundance. For the sake of simplicity, we only choose a set of typical values in the parameter space such as
\begin{alignat}{1}
 &v_{1}=2000\:\mathrm{TeV},\hspace{0.5cm} v_{2}=250\:\mathrm{GeV},\hspace{0.5cm} v_{H}=246\:\mathrm{GeV}, \nonumber\\
 &M_{\Phi}=5\:\mathrm{TeV},\hspace{0.5cm} M_{\phi_{2}^{0}}=150\:\mathrm{GeV},\hspace{0.5cm} M_{h^{0}}=125\:\mathrm{GeV}, \nonumber\\
 &M_{N_{3}}=1000\:\mathrm{TeV},\hspace{0.5cm} M_{N_{1}}=100\:\mathrm{TeV},\hspace{0.5cm} M_{\chi_{1}}=58.5\:\mathrm{GeV}, \nonumber\\
 &\lambda_{7}=10^{-7},\hspace{0.5cm} (Y_{1}^{\dagger}Y_{1})_{11}=(Y_{2}^{\dagger}Y_{2})_{11}=10^{-4}, \nonumber\\
 &(Y_{1}Y_{N}^{-1}Y_{2}^{T})_{33}=5\times10^{-4},\hspace{0.5cm} Im[(Y_{1}^{\dagger}Y_{2})_{13}(Y_{2}^{\dagger}Y_{1})_{13}]=-4.3\times10^{-7},
\end{alignat}
where I use $M_{\Phi}$ as an independent input parameter instead of $v_{\Phi}=\frac{\lambda_{7}v_{1}v_{2}v_{H}}{M_{\Phi}^{2}}\approx0.5$ MeV. All of the values in Eq. (24) are completely in accordance with the model requirements discussed in Section II. Firstly, $v_{2}$ and $M_{\chi_{1}}$ are bounded by Eqs. (20)-(22), and $M_{\phi_{2}^{0}}$ is possibly close to $M_{h^{0}}$ due to $v_{2}\sim v_{H}$. Secondly, $v_{1}, M_{\Phi},M_{N_{1}}$ are jointly bounded by Eqs. (16)-(19) as well as fitting the neutrino mass. Lastly, the Yukawa couplings are chosen as the reasonable and consistent values in view of $Y_{1}\sim Y_{2}\sim10^{-2}$. It should be stressed that we do not make any fine-tuning in Eq. (24), only $M_{\chi_{1}}$ and $Im[(Y_{1}^{\dagger}Y_{2})_{13}(Y_{2}^{\dagger}Y_{1})_{13}]$ are taken as the two precise values in order to fit $\Omega_{\chi_{1}}h^{2}$ and $\eta_{B}$ respectively, while the rest of the parameters are roughly fixed to their order of magnitudes.

  Now put Eq. (24) into the foregoing equations, we can correctly reproduce the desired results,
\ba
 m_{\nu_{3}}\approx 0.043\:\mathrm{eV},\hspace{0.5cm} \eta_{B}\approx 6.2\times10^{-10}, \hspace{0.5cm} \Omega_{\chi_{1}}h^{2}\approx 0.12\,,
\ea
they are in agreement with the current experimental data very well \cite{1}. Here we only give the upper bound of neutrino mass which is assumed as $m_{\nu_{3}}$. The full experimental data of the neutrino masses and mixing angles can completely be fitted by choosing suitable texture of the matrix $Y_{1}Y_{N}^{-1}Y_{2}^{T}$. In addition, we can calculate out $\frac{\Gamma_{tree}}{H}\approx0.07$ by Eqs. (16)-(17), this clearly demonstrates that the decay of Fig. 1 is indeed severely out-of-equilibrium.

\begin{figure}
 \centering
 \includegraphics[totalheight=8cm]{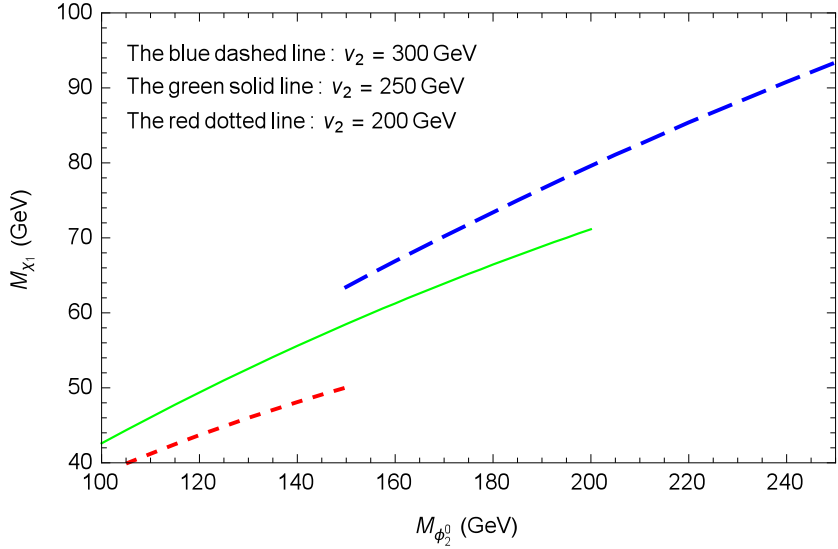}
 \caption{The curves of $M_{\phi^{0}_{2}}$ versus $M_{\chi_{1}}$ of the dark neutral scalar and the CDM $\chi_{1}$ for the three cases of $v_{2}=[300,250,200]$ GeV, each point of the curves can correctly fit $\Omega_{\chi_{1}}h^{2}\approx0.12$.}
\end{figure}
  Fig. 3 shows the three curves of $M_{\phi^{0}_{2}}$ versus $M_{\chi_{1}}$ for the three cases of $v_{2}=300$ GeV, $v_{2}=250$ GeV, and $v_{2}=200$ GeV, each point of the curves can correctly fit $\Omega_{\chi_{1}}h^{2}\approx0.12$. Note that the other parameters are not involved in this fitting. Evidently, a reasonable and moderate value of $v_{2}$ is in the range of $200\:\mathrm{GeV}\lesssim v_{2}\lesssim 300\:\mathrm{GeV}$. If $v_{2}$ is too high, the $\chi_{1}$ annihilation cross-section will be too small, then the $\chi_{1}$ relic abundance will be overclose. Conversely, if $v_{2}$ is too low, the $\chi_{1}$ annihilation cross-section will be too large, then the $\chi_{1}$ relic abundance will be deficiency. Therefore $v_{2}$ should be around the electroweak scale $v_{H}$. In addition, the curves clearly indicate that the $\phi^{0}_{_{2}}$ mass is probably in the area of $100\:\mathrm{GeV}\lesssim M_{\phi^{0}_{2}}\lesssim 250\:\mathrm{GeV}$, and the CDM $\chi_{1}$ mass is probably in the range of $40\:\mathrm{GeV}\lesssim M_{\chi_{1}}\lesssim 90\:\mathrm{GeV}$. In short, the future experimental search for $\chi_{1}$ and $\phi_{2}^{{0}}$ should focus on this parameter space of Fig. 3.

  In the end, we simply discuss the test of the model. Some new particles can be produced at the TeV-scale colliders. The relevant processes are as follows,
\begin{alignat}{1}
 &p+p \rightarrow\gamma+\gamma\rightarrow \Phi+\overline{\Phi},\hspace{0.5cm} e^{-}+e^{+}\;\mbox{or}\;p+\overline{p} \rightarrow\gamma\rightarrow \Phi+\overline{\Phi}, \nonumber\\
 &\Phi\rightarrow l_{\alpha}+l_{\beta}+\overline{H},\hspace{0.5cm} \Phi\rightarrow H+\phi_{2}^{0}\;\mbox{or}\;H+G^{0},\hspace{0.5cm} \phi_{2}^{0}\rightarrow \chi_{1}+\chi_{1}\;\mbox{or}\;G^{0}+G^{0}.
\end{alignat}
At the present LHC \cite{22}, we have a chance to search $\Phi$ and $\overline{\Phi}$ through two gamma photon fusion if the collider energy can reach their masses, but this detection is very difficult because its cross-section is too small. A better way to produce $\Phi$ and $\overline{\Phi}$ is at the $e^{-}+e^{+}$ or $p+\overline{p}$ colliders via the gamma photon s-channel mediation as long as the center-of-mass energy is enough high, for instance, the future colliders such as CEPC and ILC have some potentials to achieve this goal \cite{23}. Only if $\Phi$ and $\overline{\Phi}$ are produced, firstly, we can directly test the leptogenesis mechanism of the model by the decay asymmetry of $\Phi\rightarrow l_{\alpha}+l_{\beta}+\overline{H}$ and $\overline{\Phi}\rightarrow \overline{l_{\alpha}}+\overline{l_{\beta}}+H$. Secondly, this can indirectly shed light on the neutrino mass origin, namely the model see-saw mechanism. Thirdly, we can probe the dark particles $\phi_{2}^{0}$ and $G^{0}$ by $\Phi\rightarrow H+\phi_{2}^{0}$ and $\Phi\rightarrow H+G^{0}$. Lastly, $\phi_{2}^{0}$ can decay into a pair of the CDM $\chi_{1}$ or $G^{0}$, by which we can measure the $\chi_{1}$ mass and find the Goldstone boson. All kinds of the final state signals are very clear in the decay chain of $\Phi$ and $\overline{\Phi}$. These search are possibly a intriguing direction in the future collider experiments.

  Of course, the model can also be tested by some non-collider experiments. An indirect detection for the CDM $\chi_{1}$ is a search for the high-energy gamma photon and Goldstone boson in the cosmic rays \cite{24}, they arise from an annihilation process of the CDM $\chi_{1}$ in the dark galactic halo as shown Fig. 4.
\begin{figure}
 \centering
 \includegraphics[totalheight=5cm]{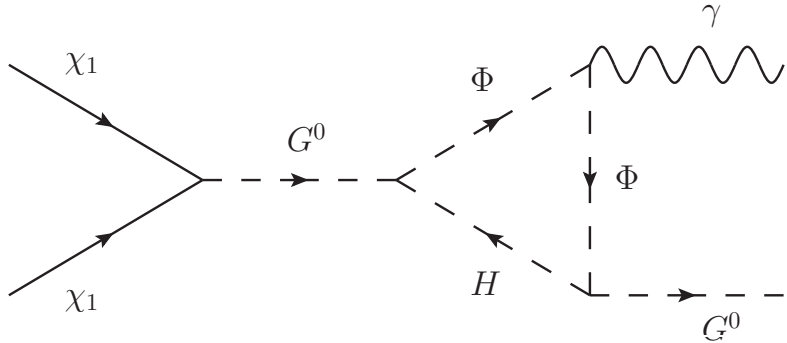}
 \caption{The indirect detection of the CDM $\chi_{1}$ through a search for the high-energy gamma photon and Goldstone boson in the cosmic rays, and its mass can accurately be measured by the energy relation $E_{\gamma}=E_{G^{0}}=M_{\chi_{1}}$.}
\end{figure}
Since both gamma photon and Goldstone boson are massless, their energy are fixed as $E_{\gamma}=E_{G^{0}}=M_{\chi_{1}}$ due to conversation of energy and momentum. If the gamma ray whose energy is $40-90$ GeV is found, this is not only a definite signal of the process of Fig. 4, but also it can tell us the accurate mass of the CDM $\chi_{1}$. A direct detection is very difficult by means of scattering off nuclei at the underground detectors such as DAMA, XENON, etc. \cite{25}, but it is not impossible. The diagram of the scattering process is similar to Fig. 4 but the photon line becomes an internal one and it is connected with a proton as an external line. In short, it will be very large challenges to actualize the above-mentioned experiments, this needs the researchers make a great deal of efforts, however, its scientific significance is beyond all doubt. We will give an in-depth discussion on the model test in another paper.

\vspace{0.6cm}
\noindent\textbf{VI. Conclusions}

\vspace{0.3cm}
  In summary, I suggest a new extension of the SM by introducing the dark sector with the local $U(1)_{D}$ symmetry. The particles in the dark sector have all non-vanishing $D$ numbers, while all of the SM particles have no $D$ numbers. The model also conserves the global $B-L$ symmetry and the hidden discrete $Z_{2}$ one. The three symmetries are together broken by $\langle\phi_{1}\rangle$ at the scale of thousands of TeVs, but the global $B-L-D$ is kept as a residual symmetry. This breaking gives rise to heavy neutral gauge boson $M_{Z'}$ and neutral Dirac fermion $M_{N}$ in the dark sector. When the universe temperature is close to the electroweak scale, the global $B-L-D$ is violated by $\langle\phi_{2}\rangle$, this generates the CDM $\chi_{1}$ mass and leads to the ``WIMP Miracle". The dark doublet scalar $\Phi$ with several TeVs mass can decay into two left-handed doublet leptons and one doublet Higgs of the SM, this process can elegantly achieve the leptogenesis at the TeV scale. The tiny neutrino mass is generated by the hybrid see-saw mechanism, it is suppressed by both the heavy $M_{N}$ and the small $\langle\Phi\rangle$ induced from the feeble scalar coupling. In brief, the model with fewer parameters is a simple and natural extension of the SM, it can collectively account for the tiny neutrino mass, the matter-antimatter asymmetry, and the CDM. In particular, the model gives some interesting predictions, for example, the leptogenesis at the TeV scale, the CDM $\chi_{1}$ with dozens GeVs mass, the dark neutral scalar boson $\phi_{2}^{0}$ with $100-250$ GeV mass, the dark background radiation of Goldstone bosons with a tiny abundance, all of them are possibly probed by the TeV collider experiments, the underground detectors, and the cosmic ray search. In short, new physics of the dark sector beyond the SM sector are very attractive and worth researching in depth.

\vspace{0.6cm}
 \noindent\textbf{Acknowledgements}

\vspace{0.3cm}
  I would like to thank my wife for her large helps. This research is supported by the Fundamental Research Funds for the Central Universities Grant No. WY2030040065.

\end{document}